\documentclass[11pt,a4paper]{article}
\usepackage{jcappub}
\usepackage{siunitx}

\newcommand{\edit}[1]{\textcolor{black}{#1}}

\usepackage{amsmath}
\graphicspath{{./}{}}

\title{Probing Ultra-light Axion Dark Matter from 21cm Tomography using Convolutional Neural Networks }

\emailAdd{kadota.kenji@f.nagoya-u.jp}

\author[a]{Cristiano G. Sabiu}
\affiliation[a]{Natural Science Research Institute, University of Seoul, 163 Seoulsiripdaero, Dongdaemun-gu, Seoul, 02504, Republic of Korea}

\author[b,c,d,e]{Kenji Kadota}
\affiliation[b]{School of Fundamental Physics and Mathematical Sciences,
Hangzhou Institute for Advanced Study, UCAS, Hangzhou 310024, China}
\affiliation[c]{International Centre for Theoretical Physics Asia-Pacific (ICTP-AP), Beijing/Hangzhou, China}
\affiliation[d]{Center for Theoretical Physics of the Universe, Institute for Basic Science (IBS), Daejeon, 34051, Republic of Korea}
\affiliation[e]{Kobayashi-Maskawa Institute for the Origin of Particles and the Universe, Nagoya University, Nagoya, 464-8602, Japan}
\author[f]{Jacobo Asorey}
\affiliation[f]{Centro de Investigaciones Energeticas, Medioambientales y Tecnologicas (CIEMAT), Av. Complutense, 40, 28040 Madrid, Spain}

\author[a,g]{Inkyu Park}
\affiliation[g]{Department of Physics, University of Seoul, 163 Seoulsiripdaero, Dongdaemun-gu, Seoul, 02504, Republic of Korea}

\abstract{
We present forecasts on the detectability of Ultra-light axion-like particles (ULAP) from future 21cm radio observations around the epoch of reionization (EoR). We show that the axion as the dominant dark matter component has a significant impact on the reionization history due to the suppression of small scale density perturbations in the early universe. This behavior depends strongly on the mass of the axion particle. 

Using numerical simulations of the brightness temperature field of neutral hydrogen over a large redshift range, we construct a suite of training data. This data is used to train a convolutional neural network that can build a connection between the spatial structures of the brightness temperature field and the input axion mass directly. We construct mock observations of the future Square Kilometer Array survey, SKA1-Low, and find that even in the presence of realistic noise and resolution constraints, the network is still able to predict the input axion mass. We find that the axion mass can be recovered over a wide mass range with a precision of approximately 20\%, and as the whole DM contribution, the axion can be detected using SKA1-Low at 68\% if the axion mass is $M_X<1.86 \times10^{-20}$eV although this can decrease to $M_X<5.25 \times10^{-21}$eV if we relax our assumptions on the astrophysical modeling by treating those astrophysical parameters as nuisance parameters.
}
\arxivnumber{2108.07972 }
\begin{document}

\maketitle
\section{Introduction} \label{sec:intro}
Despite the remarkable success of the cold dark matter model in $\Lambda$CDM cosmology, the nature of dark matter, such as its mass and interaction rate, remains unknown. Among the popular candidates is the weakly interacting massive particle (WIMP), however its lack of detection from a variety of experimental searches motivates us to explore the wider range of dark matter properties beyond, for instance, those typical in the weak scale supersymmetry (the conventional values for the thermal WIMPs are the mass of $\sim {\cal O}(10^{2-3})$ GeV and the thermally averaged annihilation cross section $\langle \sigma v\rangle \sim {\cal O}(10^{-26})cm^3/s$) \cite{Jungman:1995df}.  As an alternative to the conventional  $\Lambda$CDM, the ultra-light particle (whose mass can be as light as $\sim 10^{-21}$ eV and sometimes dubbed `fuzzy dark matter') has gained the revived interest whose wave-like nature is characterized by its de Broglie wavelength below which the fluctuation growth is suppressed due to `quantum pressure' \citep{Hu:2000ke, Hui:2016ltb, Safarzadeh:2019sre,Hui:2021tkt}. 

Such a matter fluctuation suppression at small scales is motivated from the discrepancies between the $\Lambda$CDM simulations and the observations, such as the simulation's stellar velocity dispersion larger than observed ones, even though more detailed analysis of baryonic processes would be required to study those small scale structures. Another motivation for such a light particle comes from the symmetries which could exist in the early Universe. An ultra-light particle can well be a remnant of the symmetry which was restored at a high energy scale, such as a pseudo-Nambu-Goldstone boson (pNGB) arsing from the spontaneous symmetry breaking of a continuous global symmetry. Axion is a well known example which arises due to the spontaneous breaking of Peccei-Quinn (PQ) symmetry introduced to solve the QCD strong-CP problem \citep{Peccei:1977hh,Weinberg:1977ma,Wilczek:1977pj,Preskill:1982cy,Abbott:1982af,Dine:1982ah,Sikivie:1983ip}. Axion can also serve as a promising dark matter candidate \citep{Raffelt:1987im,ADMX:2009iij,Kadota:2014hpa,Lee:2017qve,CAST:2017uph,Kadota:2013iya,Kelley:2017vaa,Huang:2018lxq,Hook:2018iia,Kadota:2015uia,Harari:1992ea,Fedderke:2019ajk,Kadota:2019ktm,Shimabukuro:2019gzu,Bak:2020det}, and, more generally, the axion-like particles where the ultra-light mass and coupling are treated independently have gained growing interests as an alternative to the conventional $\Lambda$CDM \citep[such scenarios with a wide range of axion mass e.g. from $10^{-33}$eV to $10^{-10}$eV are sometimes referred to as string axiverse; ][]{Svrcek:2006yi,Arvanitaki:2009fg,Hu:2000ke,Amendola:2005ad,Hui:2016ltb,Marsh:2015xka,Kadota:2013iya}.

We call such ultra-light scalar particles which can constitute the whole or non-negligible fraction of the dark matter the axion in this paper. The observational signals of small scale suppression such as Ly-alpha and 21cm signals have been explored which can potentially lead to the stringent bounds on the axion mass \citep[$>20\times10^{-22}$eV,][]{Irsic:2017yje}.  Such suppression can also lead to the delay of epoch of reionization (EoR), which is the focus of our paper. EoR whose details still remain to be explored is among the main targets for the forthcoming 21cm experiments such as the Square Kilometer Array (SKA) \cite{SKA:2018ckk}. 

The goal of this paper is to demonstrate the feasibility of machine learning to extract the axion particle mass from the tomographic maps of the differential brightness temperature of neutral hydrogen. Convolutional Neural Networks (CNN) in particular are well suited to image analysis and have been shown to perform well on simulated 21cm data \citep{2020JKPS...77...49K}, thus we utilize them in this work to study the spatial distribution of the 21cm signal over a wide range of redshift.
We use the \texttt{21cmFAST} code to produce the differential brightness temperature maps which constitute the training and testing data sets for our studies \citep{Mesinger:2010ne}.



In section \ref{sec:model} we explain how our 21cm maps are constructed and how the axion DM model is incorporated into this framework. We also take into account observational effects and noise associated with the future SKA design. In section \ref{sec:method}, we introduce our machine learning methodology, define the specific architecture for our convolutional neural network and design our training and test data sets.  Section \ref{sec:results} shows the ability of our machine learning algorithm to constrain the axion mass parameter and to differentiate our axion model (where the axion constitutes the entirety of dark matter) and the conventional $\Lambda$CDM model without axion.

Throughout this paper and unless otherwise stated the fiducial  cosmological model is chosen to be flat $\Lambda$CDM, with $\Omega_M=0.315$, $\Omega_b=0.049$, $h=0.67$, and $\sigma_8=0.829$ corresponding to the best fit Planck 2015 measurements \citep{Planck:2015fie}.

\section{Modeling the 21cm signal} \label{sec:model}
Neutral hydrogen in its ground state has a hyperfine splitting 
which corresponds a frequency of 1420 MHz. The relative occupations of these two levels is determined by the spin temperature $T_s$ through,
\begin{equation}
    \frac{n_1}{n_0}=3e^{-\frac{T_{\star}}{T_{s}}},
\end{equation}
where $T_{\star}=0.068$K is the temperature corresponding to the energy difference between these two levels and $n_1$ and $n_0$ are the number densities of neutral hydrogen in the excited and ground states respectively. 

The frequency of the redshifted 21cm signal lies in the radio range and within the Rayleigh Jean’s limit the intensity of this spectral line can be quantified by the brightness temperature $T_b$. This is observed as an offset from the CMB temperature and is called the differential brightness temperature $\delta T_b$. For an observed frequency $\nu$, corresponding to a redshift $z$ at a given point in space $x$ \citep{Furlanetto:2006jb},
\begin{equation}
       \delta T_b(\nu,x)\approx 27x_H(x)(1+\delta(x))\left( 1-\frac{T_{\gamma}(z)}{T_s(x)}\right)\frac{\Omega_bh^2}{0.023}\left(\frac{1+z}{10}\frac{0.15}{\Omega_Mh^2}\right)^{1/2} \text{[mK]}
\end{equation}
where $x_H$ is the fraction of neutral hydrogen, $\delta=\bar{\rho}/\rho -1$ is the over-density relative to the mean density $\bar{\rho}$, and $T_{\gamma}$ is the CMB temperature.

\subsection{21cm signal}
We simulate the 21-cm maps using the semi-numerical  simulation code \texttt{21cmFAST} \citep{Mesinger:2007pd,Mesinger:2010ne}. This code generates a Gaussian initial density field from an input power spectrum $P_k$ and evolves it according to first order perturbation theory. It generates the density $\delta$, spin temperature $T_s$,  the fraction of neutral hydrogen $x_H$ and the differential brightness temperature $\delta T_b$ at every grid point within the simulation box at a given redshift. It can also output the data in a lightcone format where one axis acts as a redshift dimension. 

Within \texttt{21cmFAST} there are several astrophysical parameters that affect $\delta T_b$ across the reionisation epoch, defined when the first energetic sources begin to ionize neutral hydrogen until we are left with a fully ionized plasma.  The rate of production of UV photons is determined by the ionization  efficiency  of  star-forming  galaxies, which depends on the fraction of UV photons that escape the galaxy, $f_{esc}$, the fraction of gas in stars $f_{star}$, the  number of ionizing photons per stellar baryon $N_{\gamma}$ and the number of times a hydrogen atom recombines on average $n_{rec}$.
However, within 21cmFAST these 4 variables are encapsulated within a single parameter $\zeta$, 
\begin{equation}
    \zeta = 30 \left(\frac{f_{esc}}{0.15}\right)\left(\frac{f_{\star}}{0.1}\right)\left(\frac{N_{\gamma}}{4000}\right)\left(\frac{2}{1+n_{rec}}\right).
    \label{eq:zeta}
\end{equation}

Within the code ionized regions are chosen when $\zeta f_{coll}(x,z) \geq 1$ \citep{Furlanetto:2004nh}, where $f_{coll}$ is the collapse fraction and can be computed using the extended Press-Schechter model \citep{Bond:1990iw,Lacey:1993iv}. In this work we will set $\zeta=30$ as a fiducial value however in \S\ref{sec:results2} we will relax this assumption and show how uncertainties on this quantity will impact our results.


Another important parameter is the minimum virial temperature of halos producing ionizing  photons during the EoR. Typically, $T_{vir}$ is chosen to be $10^4$K, corresponding to the temperature above which atomic cooling becomes effective 
to trigger gravo-thermal instability and thus star formation. Throughout this work we adopt $T_{vir}=10^4$K as a fiducial value, although in \S\ref{sec:results2} we will investigate a range of values to quantify its effect on our estimation of axion mass.
\edit{The mean free path of ionizing photons, $R_{mfp}$, is another astrophysical parameter of \texttt{21cmFAST}. In this work we kept it fixed at its default value of 15Mpc.}

\subsection{Axion Dark Matter}
\begin{figure}
    \centering
    \includegraphics[width=0.6\columnwidth]{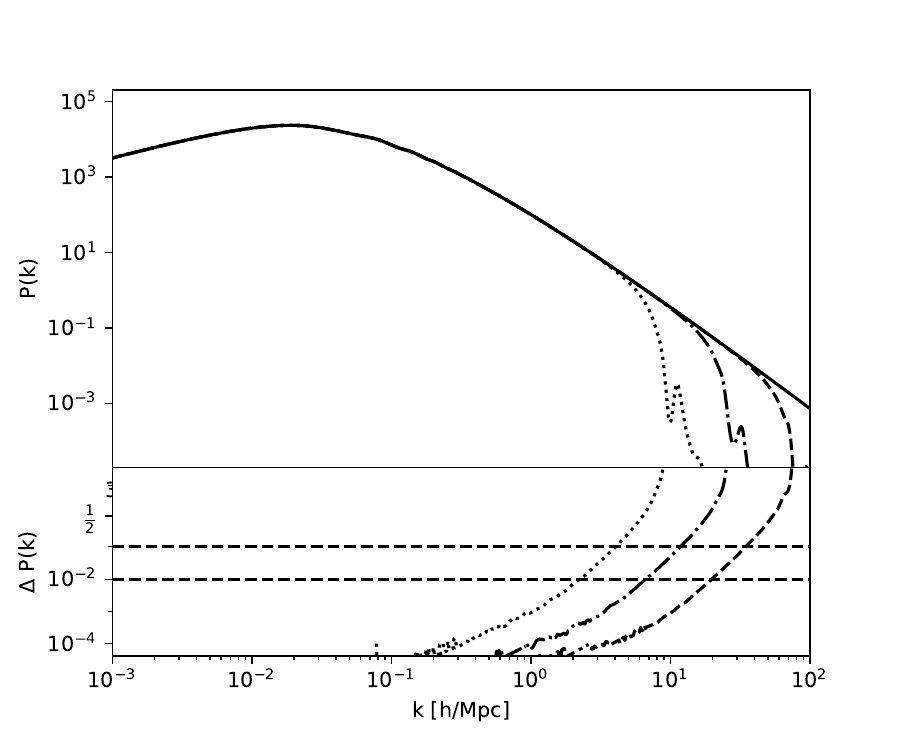}
    \caption{The linear matter power spectrum at z=0 for $\Lambda$CDM without axion (black solid line) and for axion scenarios with various axion masses: $M_X=10^{-20}$eV (dashed line), $M_X=10^{-21}$eV (dot-dashed) and $M_X=10^{-22}$eV (dotted). In the lower panel we display the fractional difference in the power spectra for each axion model compared to  $\Lambda$CDM with a 1\% and 10\% reduction shown as the horizontal black dashed lines.
    }
    \label{fig:axion_pk}
\end{figure}

We generate linear matter power spectra for axion dark matter models using the \texttt{AxionCAMB}\footnote{\url{https://github.com/dgrin1/axionCAMB}} code  a modified version of the cosmological Boltzmann code, \texttt{CAMB}. Assuming our fiducial cosmological model parameters we can see the effect of varying the axion particle mass on the power spectrum in figure~\ref{fig:axion_pk}. The suppression of power at large $k$ will have the effect of smoothing small scale structures. For lighter masses this effect is more prominent but as the mass increases the power spectrum becomes almost indistinguishable from the standard $\Lambda$CDM model over the scales shown in this figure. 
The suppression is expected to show up for the scales smaller than the de Broglie wavelength, which is inversely proportional to the momentum, $\lambda =2\pi /p \sim 0.1[kpc] (10^{-21}[eV]/M_X) (10^2[km/s] /v)$.

\begin{figure}
    \centering
     \includegraphics[width=0.65\columnwidth]{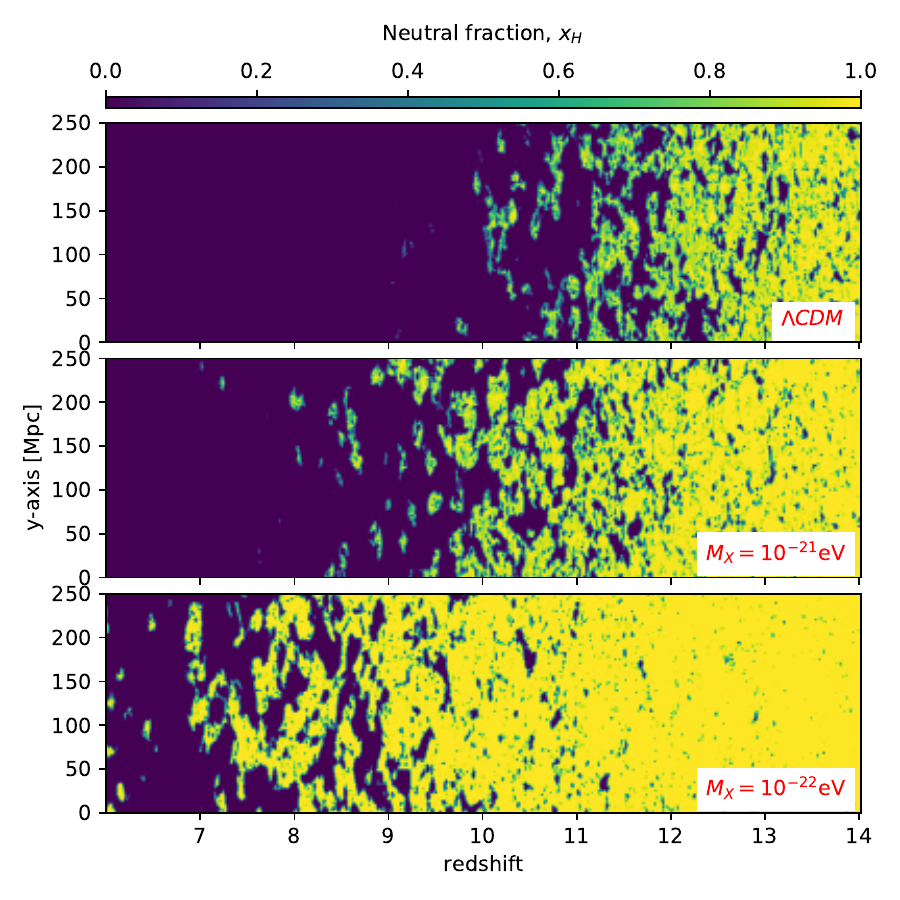}
     \includegraphics[trim={0.5cm 0 0.4cm 0},clip,width=0.33\columnwidth]{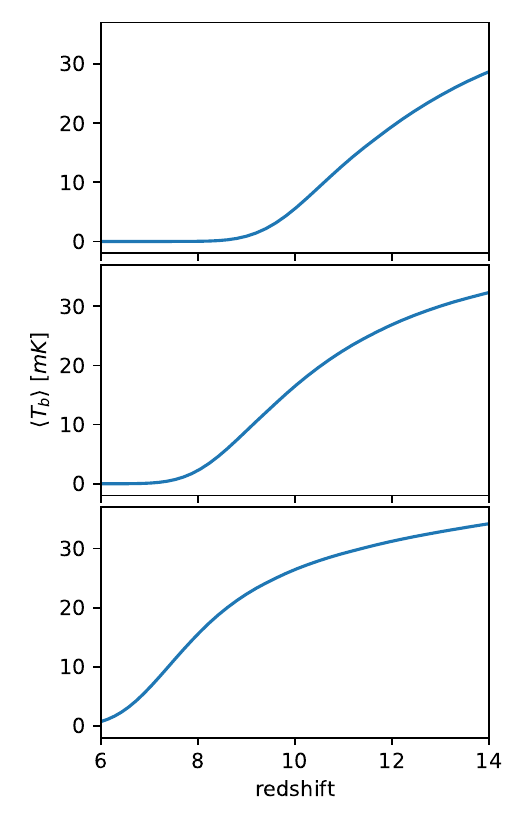}
    \caption{{\em Left:} Simulated spatial distribution of the fraction of neutral hydrogen over the redshift range $6<z<14$ for three models. \edit{{\em Right:} The global brightness temperature, $\left<T_b \right>$, as a function of redshift for three models.} In the top panels we show the $\Lambda$CDM case while the middle and bottom panels show axion models with particle masses of $10^{-21}$ and $10^{-22}$eV respectively. The fiducial model is flat $\Lambda$CDM, with $\Omega_M=0.315$, $\Omega_b=0.049$, $h=0.67$, $\sigma_8=0.829$, $T_{vir}=10^4$K and $\zeta=30$.}
    \label{fig:lightcones}
\end{figure}

\begin{figure}
    \centering
    \includegraphics[width=0.6\columnwidth]{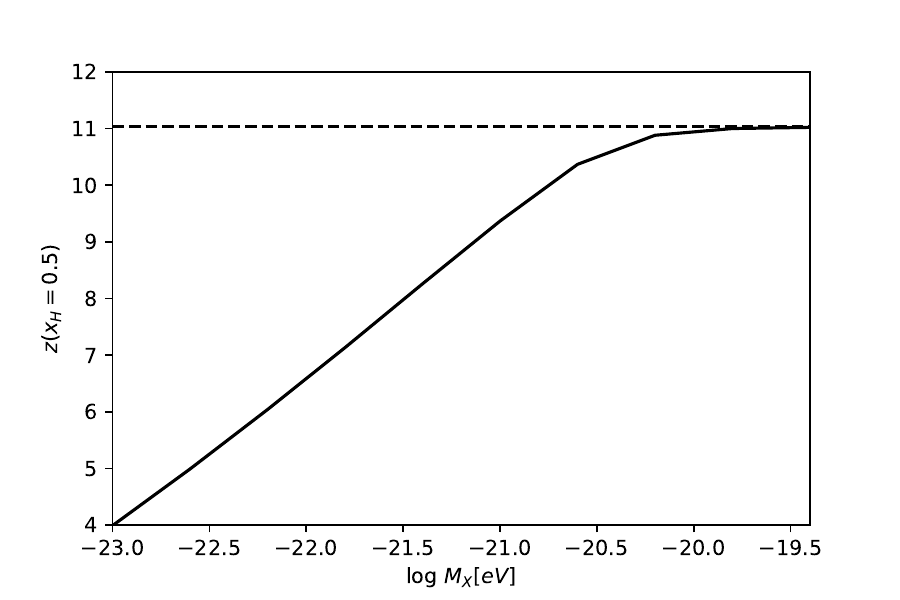}
    \caption{The redshift of reionization (i.e. when the volume averaged $\langle x_H\rangle=0.5$) as a function of axion mass (Black line). The redshift of reionization for the $\Lambda$CDM without axion is shown as dashed line.
    }
    \label{fig:reionize}
\end{figure}

We now proceed to run \texttt{21cmFAST} with axion DM power spectra.
In figure~\ref{fig:lightcones} we see the effect that the axion mass has on the formation of ionized regions. Lighter axion masses suppress the growth of ionized `bubbles', delaying reionization, compared to higher axion masses and to standard $\Lambda$CDM. 
This is due to the power spectrum suppression in the presence of axions as can be seen in figure \ref{fig:axion_pk}.
The suppression of density fluctuations has a significant impact on the collapse fraction of gas \citep{Sitwell:2013fpa,Sarkar:2015dib,Nebrin:2018vqt,Jones:2021mrs} and pushes the reionization to lower redshift. If we measure the reionization redshift, $z_{reion}$ as the redshift at which half of the hydrogen becomes ionized we can plot what this looks like as a function solely of the axion mass. In figure~\ref{fig:reionize} we show $z_{reion}$ as a function of the axion mass and see that at larger masses ($\approx 10^{-19.5}$eV) the reionization redshift is very close to that of$\Lambda$CDM (dashed line).

For lighter masses this ionization redshift is pushed to much lower redshift and at $M_X=10^{-23}$eV $z_{reion}\approx 4$. We consider the axion mass range of $10^{-23} - 10^{-19}$eV in the following numerical analysis for the illustration purpose, even though those small axion masses $M_X\lesssim10^{-22.5}eV$ are excluded because the reionization should be complete by $z\sim 6$ \citep{SDSS:2001tew}.

\subsection{SKA1-low mock observations}
The full width at half-maximum (FWHM) of the point spread function (PSF) of an interferometric radio system can be written as,
\begin{equation}
    \Delta\theta=\frac{\lambda}{B},
    \label{eq:resolution}
\end{equation}
where B is the maximum baseline and $\lambda$ is the wavelength.
For our interest in observing the redshifted 21cm signal the above equation acquires a redshift dependence with $\lambda=\lambda_{21}(1+z)$. In this work we adopt the SKA1-Low design, which will have a central core of antennae with radius $\sim 500$m. 

Thus $\Delta\theta$ varies from $\sim$\SI{7.2}{\arcminute}, corresponding to a comoving angular diameter distance of 3.1Mpc/$h$, at z=4.0, to \SI{21.8}{\arcminute}, with a scale of 4.4Mpc/$h$, at z=14. We use this varying resolution to smooth our simulated maps in angular space at each redshift/frequency slice.

In order to simulate realistic observations, we also consider the instrumental (thermal) noise and its impact on the observed 21cm signal. Following \citet{Alonso:2014dhk}  we add onto our simulated maps a Gaussian random field with a rms variation of,
\begin{equation}
    \sigma_{noise}=T_{sys}\sqrt{\frac{4\pi f_{sky}}{\Omega_{beam}N_{dish} t_{int}\Delta\nu}}.
    \label{eq:noise}
\end{equation}
 
where $N_{dish}$ is the number of antenna, $\Omega_{beam}=1.133\theta_{beam}^2$ is the beam size, $t_{int}$ is the integration time taken to be 1,000 hours, $\Delta\nu$ is the frequency bandwidth, which is $\sim$1MHz, $f_{sky}=0.02$ is the fraction of sky observed and the system temperature is $T_{sys}=T_{rx}+T_{gal}$ is the combination of the receiver temperature  $T_{rx}=0.1T_{gal}+40$K and the contribution from our own galaxy at frequency $\nu$, which is modelled as, $T_{gal}=25\left(\frac{\nu}{408\rm{MHz}} \right)^{-2.75}$K \citep[i.e. the SKA Red book]{SKA:2018ckk}. 

The observed 21cm signal includes a type of contamination known as the foreground emission, mostly dominated by synchrotron emission from our own galaxy. This type of contamination is usually removed following different techniques \citep{2021arXiv210710814S}. As the foreground removal usually affects more the larger scale modes of the temperature maps \citep{2020MNRAS.495.1788A,2020MNRAS.499.4613S}, we leave this for a future study and  assume for this work a perfect foreground removal.  

\begin{figure*}
    \centering
    \includegraphics[width=\columnwidth]{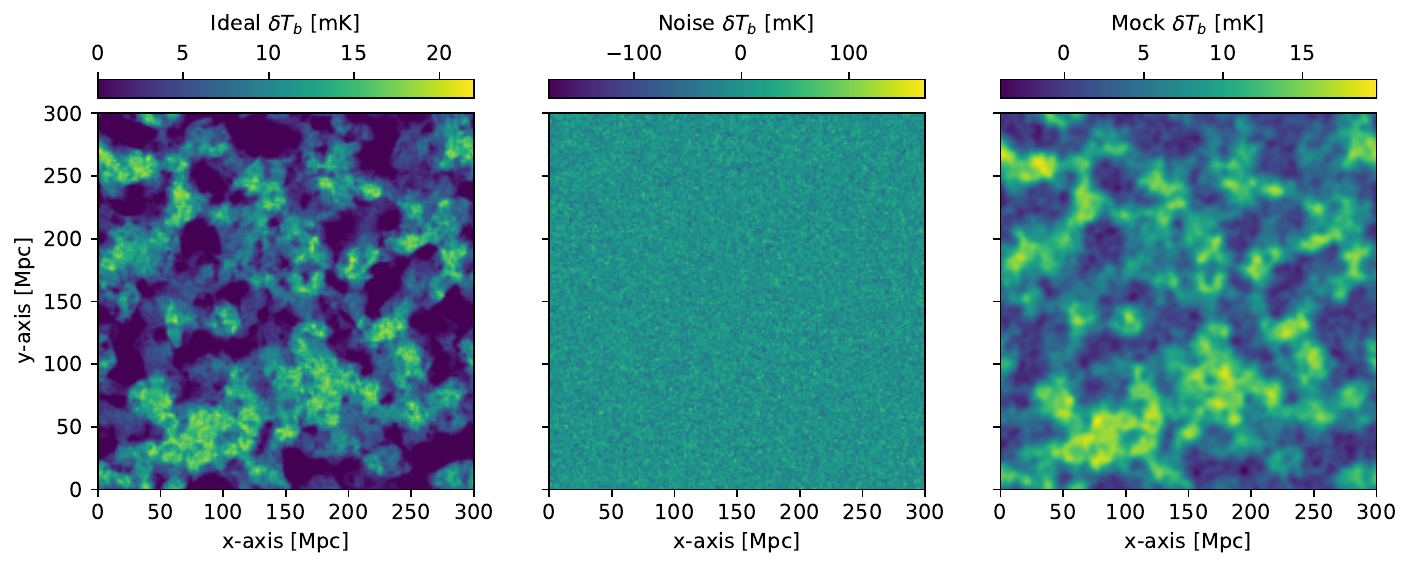}
    \caption{{\em Left: }The simulated 21cm differential brightness temperature at z=10 in a comoving slice of 30Mpc. {\em Middle: } A realisation of a noise map. {\em Right:} Same as left plot with noise and smoothing applied.}
    \label{fig:noisemap}
\end{figure*}

In figure~\ref{fig:noisemap} we show a noiseless `ideal' map output at $z=10$ in a 30cMpc slice (left panel). We then create a Gaussian noise realization corresponding to this design (middle panel), which does not display any spatial structure. In the right hand panel of figure~\ref{fig:noisemap} we see the effect of adding the noise to the ideal image and then smoothing the map with a Gaussian filter with a size corresponding to eq.~\ref{eq:resolution}. We will use this procedure to construct realistic observations in the next section.

\section{Methodology}
\label{sec:method}
The methodology that we adopt is a forward modeling approach using simulations and machine learning to regress to the parameters of interest. 

\subsection{Simulation set}
We create 400 simulations with axion masses (sampled uniformly in log-space) in the range $10^{-23} - 10^{-19}$eV while keeping the cosmological parameters fixed at the Planck15 fiducial values. \edit{Each simulation is generated from a unique Gaussian random realisation of the input power spectra, and thus our training and test data do not share similar spatial structures. This is crucial in general regression problems using neural networks as we do not want the network to learn specific localised features of the fields.} We chose the upper cutoff in the mass range considering that the difference in the power spectrum for $M_X\sim10^{-19}$eV and $\Lambda$CDM is almost indistinguishable even at scales of $k\approx100h/$Mpc. The lower mass was chosen more arbitrarily, although we note from figure \ref{fig:reionize} that the redshift of reionization for  $M_X\sim10^{-23}$eV is so low that this model would have already been ruled out by observations of the Gunn-Peterson effect in high redshift Quasars \citep{SDSS:2001tew}. We mention also that axion dark matter with masses in the range $10^{-33}$eV$\leq{M_X}\leq 10^{-24}$eV are excluded by the Planck CMB data \citep{Hlozek:2014lca,Hlozek:2017zzf}, while Lyman-alpha forest data excludes $M_X <  2\times 10^{-21}$eV \citep{Irsic:2017yje}.


The simulations are run with a box size of 300Mpc and a grid size of 256. We output the simulation in a lightcone which has 1,500 tomographic bins equally spaced in comoving distance between z=4 and 14. We reinterpret the redshift dimension as frequency and downgrade the resolution to 128 bins of width equivalent to $\Delta\nu=1.4$MHz by taking the mean of the pixel values. We then downgrade the angular dimensions to 64x64 by averaging the pixels.

We also create a set of realistic noisy simulations by applying the noise and Gaussian smoothing from Eqs~\ref{eq:noise} \& \ref{eq:resolution}, respectively, before we downgrade the dimensions of the data cube.

\subsection{Machine Learning Approach}
We implement a machine learning approach known as convolution neural networks (CNN) to build a connection between the spatial distribution of the simulated 21cm signal and the input parameters of the model. 

\begin{figure*}
    \centering
    \includegraphics[trim={0 3.5cm 0 3.5cm},clip,width=\columnwidth]{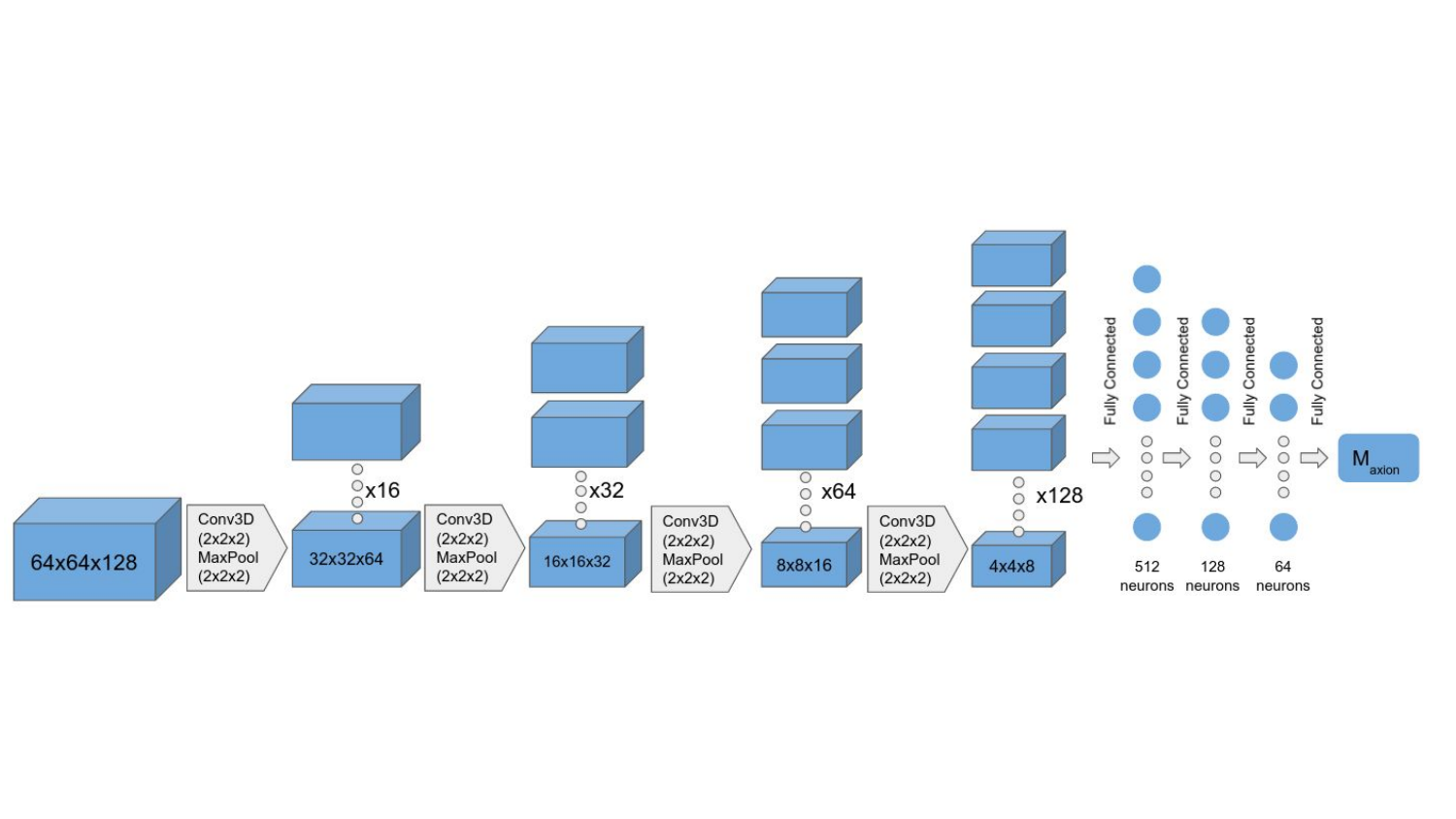}
    \caption{The architecture of our convolutional neural network.  The input data cube has 64x64x128 voxels. The four 3-dimensional convolution layers have 16,  32,  64 and 128 filters,  respectively, each using a 2x2x2 filter size and a stride of 1x1x1.  Beside each convolution layer,  a batch normalization layer is added to normalize the distribution (enhancing the stability) we then add a max pooling layer to decrease the size of the output. After the four convolution layers the elements are `flattened' to a 1-d vector, and passed to three dense fully connected neural network layers with 512, 128, 64 neurons, and finally a single output prediction for the axion mass. Between each fully connected layer a dropout layer is added to randomly zero 15\% of the weights during training.
    }
    \label{fig:cnn}
\end{figure*}

In figure \ref{fig:cnn} we show a schematic of our CNN model architecture. It takes as input a 64x64x128 lightcone data cube where each voxel has dimension 4.68cMpc$\times$4.68cMpc$\times$1.48Mhz ($cMpc$ means co-moving Mpc). The input is 3-dimensioanlly convolved with 16 (2x2x2 sized) filter then pooled down to half the spatial dimension using the maximum value in (2x2x2) voxels. This process is repeated three more times with 32, 64 and 128 filters. Before each convolution the feature maps are batch normalised to stabilise the distribution. After the convolution layers the data structure has shape 4x4x8x128 voxels, these are then flattened into a 1D vector and passed to a conventional neural network that has 3 fully connected layers of 512, 128 and 64 neurons respectively. The final layer has a single output representing our model parameter, $M_{X}$. \edit{Between each fully connected layer a dropout procedure is added which randomly zeros 15\% of the weights during training,  preventing over-fitting and is an efficient way of performing model averaging within neural networks.} Our CNN model is built using \texttt{Tensorflow/KERAS}.\footnote{\url{https://www.tensorflow.org/}} \citep{tensorflow2015-whitepaper}

We split our simulations into 3 samples: 80\% for training the network, 10\% for validating the performance of the network (but which are not used directly to tune the network weights), and 10\% for testing the fully trained network. 

We perform data augmentation and expand the training set by a factor of two by introducing random flips and rotations of $\pm90$ and 180  degrees in the 2 angular dimensions, while keeping the frequency/redshift axis unchanged. This allows the network to learn a more generalised spatial representation of the data.

\begin{figure*}
    \centering
    \includegraphics[width=0.45\columnwidth]{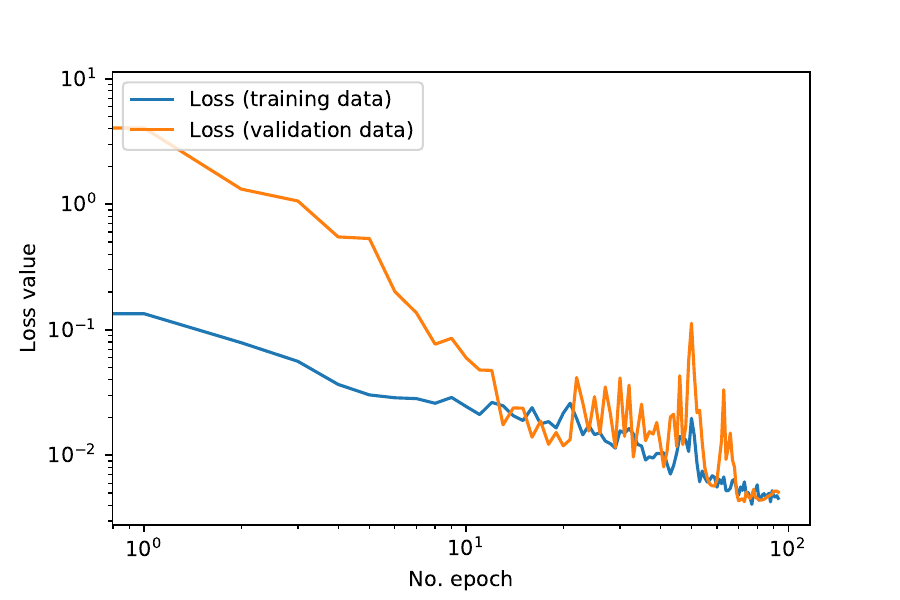}
    \includegraphics[width=0.45\columnwidth]{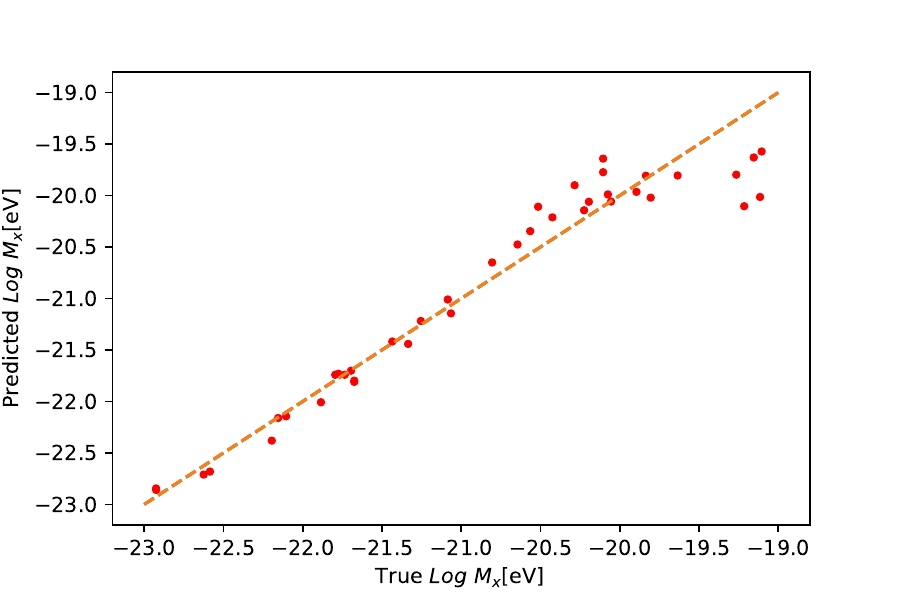}
    \caption{{\em Left:} The training MSE loss metric of our network over 100 epochs. The blue line denotes the loss of the training data while the orange line shows the loss computed on the validation data. {\em Right:} The network prediction on the testing data vs the input truth are shown as red points and the dashed line shows the one-to-one, target prediction.}
    \label{fig:CNNperformance}
\end{figure*}

In figure \ref{fig:CNNperformance} we see the training performance of the network in the left panel. The training and validation training loss track each other fairly well after 10 epochs. This reassures us that the network is not over-fitting and not learning only the specific training data.

In the right hand panel of figure \ref{fig:CNNperformance} we see the predictions of the trained network performance on the test data which has not been previously seen by the network. The predictions for $M_X$ follow closely to the one-to-one line which give us confidence that the network has learned a generalised mapping between the 3-dimensional brightness temperature field and the axion mass, under our model assumptions. It should be noted that the predicted $M_X$ matches the input truth over a wide range in axion mass, but begins to deviate when the true axion mass $M_X \gtrsim 10^{-20}$eV. The location of this plateau may be interpreted as the limit of the CNN to differentiate between the mock observations with axion masses above this mass scale. This is expected because the such a large axion mass leads to the suppression at a scale smaller than the experimentally observable scale. The limitation of CNN is also illustrated in the following section.

\section{Results} \label{sec:results}
\subsection{Axion mass prediction}
In the previous section we trained a CNN to predict the axion mass from a realistic mock observations with the SKA1-Low 21cm survey (with and without noise and smoothing). 

We now proceed to determine the accuracy of these predictions and test the ability of the network to differentiate the conventional $\Lambda$CDM from axion dark matter models. We create 100 simulation each for 4 scenarios: a conventional $\Lambda$CDM scenario and an axion dark matter scenario with $M_X=10^{-21}$eV, both with and without noise+smoothing.



\begin{figure}
    \centering
    \includegraphics[trim={0cm 2.5cm 0cm 2cm},clip,width=0.7\columnwidth]{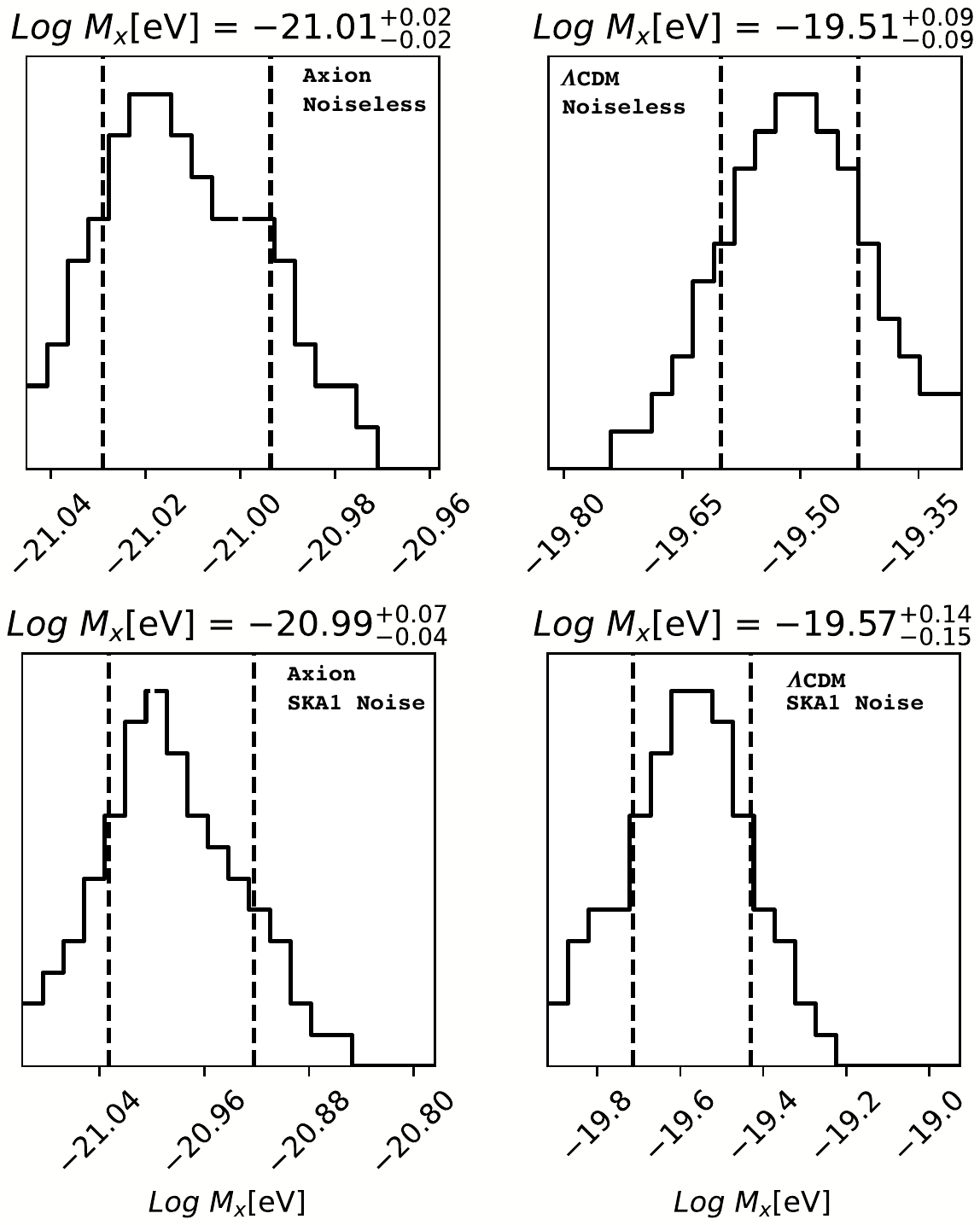}
    \caption{{\em Top Left:} The distribution of our CNN's predictions on 100 noiseless simulations fixed axion mass of $10^{-21}$eV (denoted by the vertical blue line). The vertical dashed lines show the 1-$\sigma$ (68\%) credible interval. {\em Top Right:} Same as the left plot but the simulations are created using a conventional $\Lambda$CDM scenario without axions, for which the CNN could predict the axion mass because the CNN is trained only with the data including axions.
    {\em Bottom:} As above but now each simulation has a random Gaussian noise realisation added to its signal and it is convolved with a smoothing length applicable to SKA1-Low core design computed through eq.~\ref{eq:resolution}}
    \label{fig:CNNdist}
\end{figure}

In the top panels of figure~\ref{fig:CNNdist} we show the distribution of $M_X$ predictions for the $M_X=10^{-21}$eV simulations on the left and for $\Lambda$CDM without axions on the right, under noiseless/ideal conditions. We find that the recovered axion mass for the $M_X=10^{-21}$eV simulations is $M_X=0.98^{+0.04}_{-0.05}\times10^{-21}$eV, while the  $\Lambda$CDM simulations have an inferred axion mass of $M_X=3.09^{+0.71}_{-0.58}\times10^{-20}$eV. The network was trained exclusively with simulations that include axions, therefore the inferred axion mass applying the trained CNN to the $\Lambda$CDM data without axions corresponds to the upper mass limit of the CNN's ability to detect axion dark matter.

In the bottom panels of figure~\ref{fig:CNNdist} we show similar results but now under realistic noise conditions. As we might expect the distribution is broadened and the inferred uncertainty is increased. In the bottom left we see that the mass of the axion is estimated to be $M_X=1.02^{+0.18}_{-0.09}\times10^{-21}$eV, which again recovers the true input mass. In the bottom right we see that the $\Lambda$CDM simulations are estimated to have an axion mass of $M_X=2.69^{+1.02}_{-0.78}\times10^{-20}$eV (again this is the mass range for which the CNN cannot differentiate between axion models and the conventional $\Lambda$CDM). 

We can infer from this that the axion as the whole dark matter contribution can be detected using SKA1-Low at 68\% if the axion mass is $M_X<1.86 \times10^{-20}$eV. 
Here we assume that the distribution of predicted axion masses of the standard $\Lambda$CDM simulations represents the null hypothesis and a one-tailed (left) test encapsulates the significance of the axion detection.

We can also compare the optical depth from these models with the value derived from Planck CMB data. The Planck 2018
measurement of the electron scattering optical depth is $\tau_e=0.0522\pm0.0080$ \citep[specifically TT+LowE,][]{Planck:2018vyg} which is consistent with our simulated models that give $\tau_e=0.051$ for standard $\Lambda$CDM and $\tau_e=0.041$ for an axion model with mass $M_X=10^{-21}$eV, corresponding to deviations of 0.15$\sigma$ and 1.4$\sigma$ respectively.

However it is important to note that our results so far have  assumed a fiducial cosmological model and specific choices of astrophysical parameters $T_{vir}$ and $\zeta$. While Planck and other data sets provide a tight prior which supports our assumed cosmological model, we may worry that our assumed values of $T_{vir}$ and $\zeta$, which are not well understood,  may bias our recovered axion masses as discussed in the following section.

\subsection{Parameter Degeneracies}
\label{sec:results2}
We now relax our assumptions of the astrophysical parameters and check for possible degeneracies between the predicted axion mass, $T_{vir}$ and $\zeta$.

\begin{figure*}
    \centering
    \includegraphics[trim={0 0 0.4cm 0},clip,width=0.5\columnwidth]{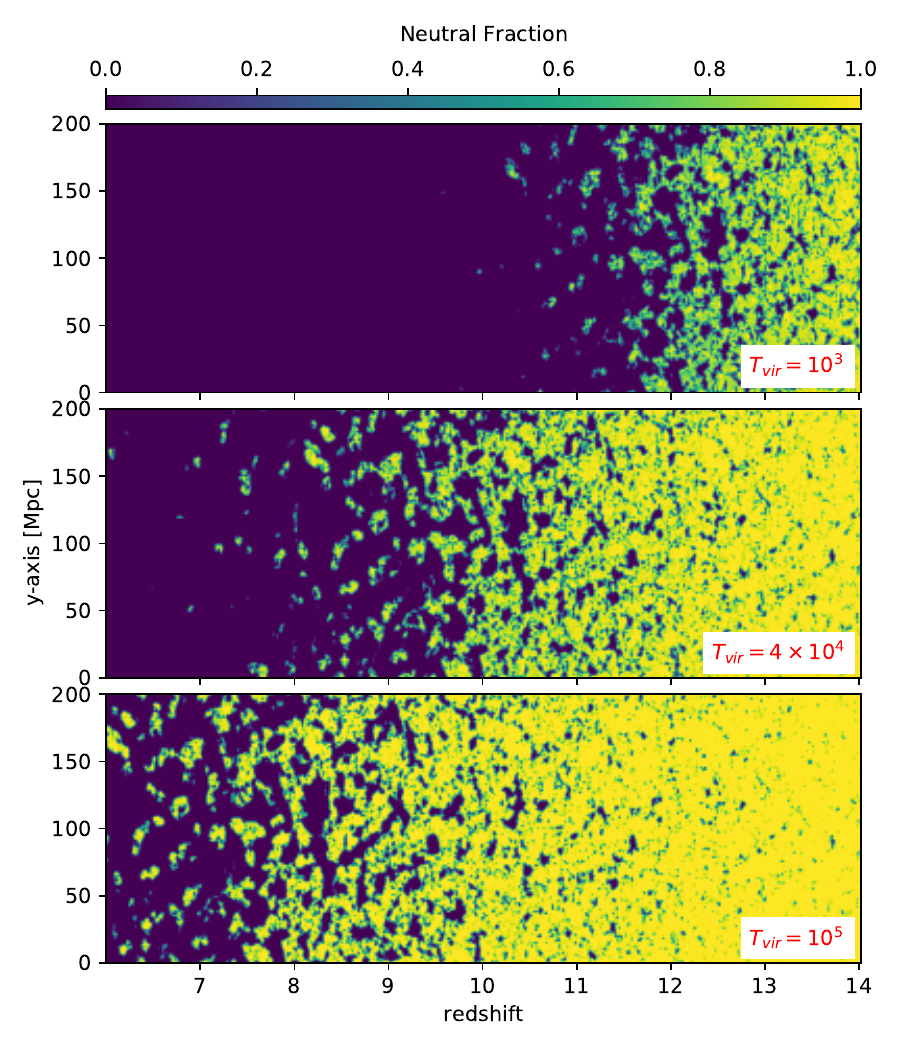}
    \includegraphics[trim={1.55cm 0 0.4cm 0},clip,width=0.448\columnwidth]{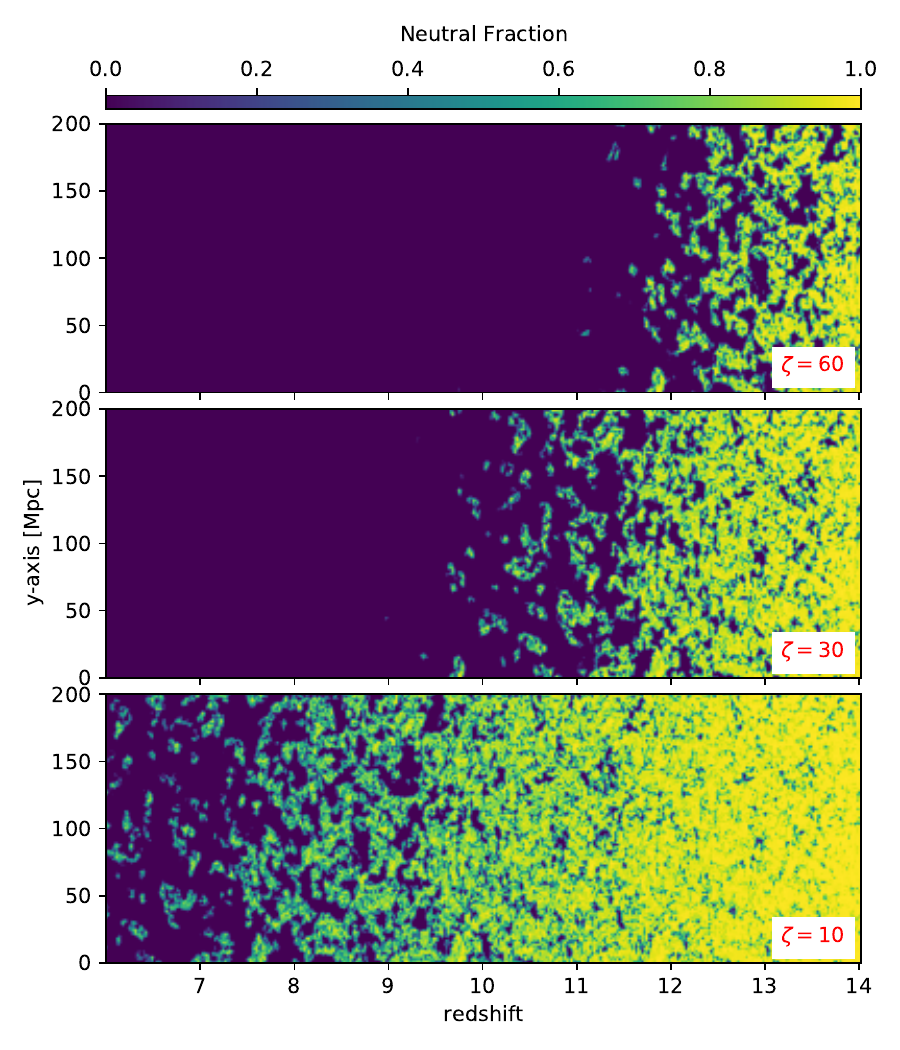}
    \caption{Simulated spatial distribution of the fraction of neutral hydrogen over the redshift range $6<z<14$ for various models. Unless otherwise stated the fiducial model is flat $\Lambda$CDM, with $\Omega_M=0.315$, $\Omega_b=0.049$, $h=0.67$, $\sigma_8=0.829$, $T_{vir}=10^4$K and $\zeta=30$. {\em Left column:} The parameter $T_{vir}$ is varied from the fiducial model with values of $10^3,4\times10^4,10^5$K from top to bottom respectively. {\em Right column:} The parameter $\zeta$ is varied from the fiducial model with values of 60, 30, and 10 from top to bottom respectively. }
    \label{fig:lightcones2}
\end{figure*}

In the left panels of figure~\ref{fig:lightcones2} we see the effect of varying $T_{vir}$ on the spatial distribution of the neutral fraction of hydrogen over a wide redshift range of $6<z<14$. Higher $T_{vir}$ causes the reionization to occur at lower redshift because the  halo masses corresponding to that temperature require additional time to grow.

In the right panels of figure~\ref{fig:lightcones2} we investigate the sensitivity of $\zeta$ on the neutral fraction. We see that for larger $\zeta$ the reionization starts earlier which is not too surprising given the nature of eq.~\ref{eq:zeta} since most of its terms contribute to the heating of the intergalactic medium. Conversely decreasing the ionizing efficiency will prolong and delay the reionization. 
It is important to note from figure~\ref{fig:lightcones2} and  figure~\ref{fig:lightcones} that the effect of varying the axion mass has a visually similar effect to varying either of the 2 astrophysical parameters, $T_{vir}$ and $\zeta$. This could pose a problem for constraining the axion mass if this degeneracy is significant.

\begin{figure}
    \centering
    \includegraphics[width=0.7\columnwidth]{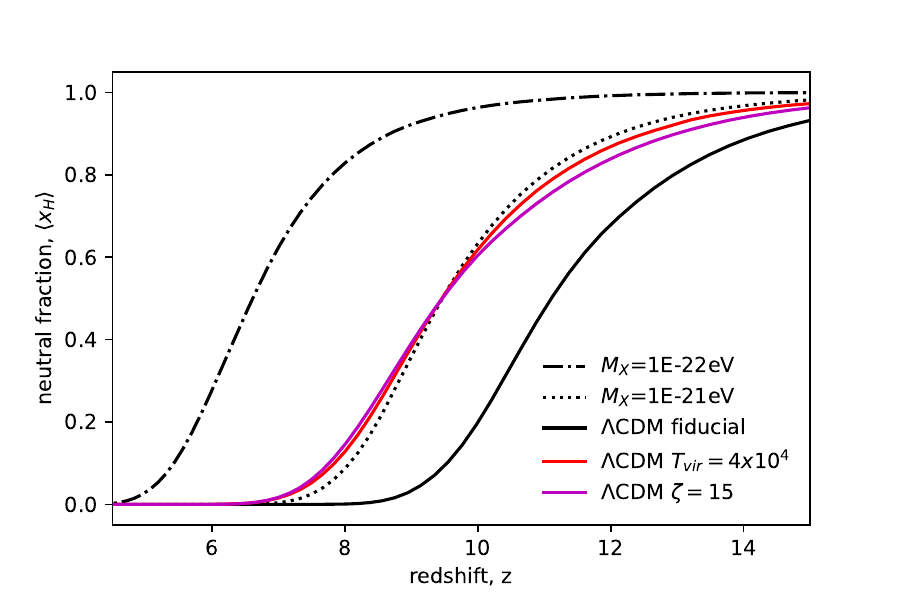}
    \caption{The Neutral fraction as a function of redshift for a standard $\Lambda$CDM model with fiducial values of $T_{vir}=10^4K$ and $\zeta=30$ (solid black) and two ultra light axion DM models with $M_X=10^{-21}$eV (dotted) and $M_X=10^{-22}$eV (dot-dashed). We also show two alternative $\Lambda$CDM models without axions i) with a high value of $T_{vir}=4\times10^4K$ (solid red) and ii) with a low value of $\zeta=15$ (solid magenta), these model values were chosen so as to match the reionization redshift of the $M_X=10^{-21}$eV axion model.}
    \label{fig:reionize2}
\end{figure}

The suppression of density fluctuations caused by axions pushes the reionization to lower redshift. We can see this clearly in figure \ref{fig:reionize2}, where we show the redshift evolution of the neutral fraction in two axion models and in the standard $\Lambda$CDM model. If we set $T_{vir}=4\times10^4$K in the $\Lambda$CDM model so that the reionization redshift matches that of the axion model with $M_X=10^{-21}$eV and $T_{vir}=10^4$K, we see that their reionization histories  track each other very closely, but not exactly. The higher $T_{vir}$ $\Lambda$CDM model has lower $x_H$ at high redshift and higher $x_H$ at lower redshift compared to the axion model. Likewise we can find the appropriate value of $\zeta(=15)$ in the fiducial $\Lambda$CDM model where its reionization redshift matches that of the axion $M_X=10^{-21}$eV model. Once again its neutral fraction follows very closely to each other but there is a clear difference which evolves with redshift indicating that tomographic observations may be able to break the degeneracy that exists between these parameters.
It is worth noting from figure \ref{fig:reionize2} that the rate of change of reionization is different depending on the value of $\zeta$, with smaller $\zeta$ prolonging the reionization epoch. On the other hand $T_{vir}$ has negligible effect on the width of the transition but it shifts the redshift. We can also see that the deviation between axion and standard $\Lambda$CDM is larger at higher redshift while they abruptly converge at lower redshift as the universe completes its reionization. This is because axion DM will suppress small scale structures which are formed at higher redshift, whereas at lower redshift there are more larger structures forming so axion suppression is less important.

\begin{figure*}
    \centering
    \includegraphics[trim={0 5.3cm 0.0cm 5.3cm},clip,width=0.7\columnwidth]{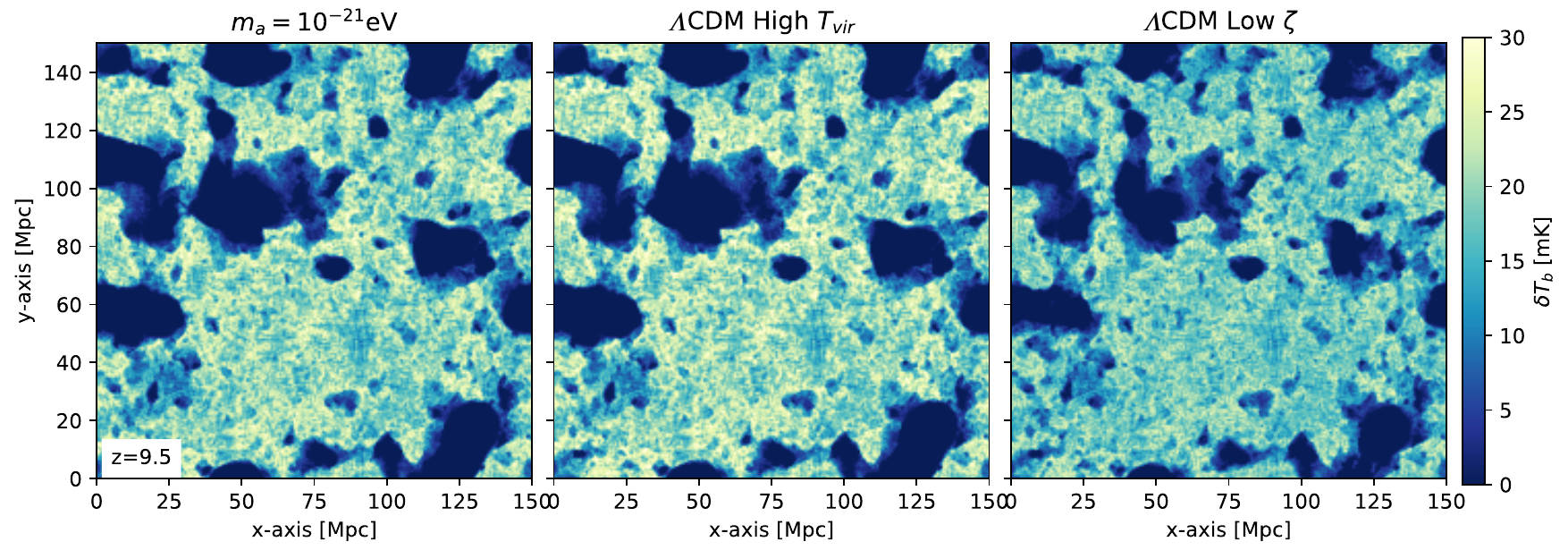}
    \includegraphics[trim={0 4.5cm 0.0cm 4cm},clip,width=0.26\columnwidth]{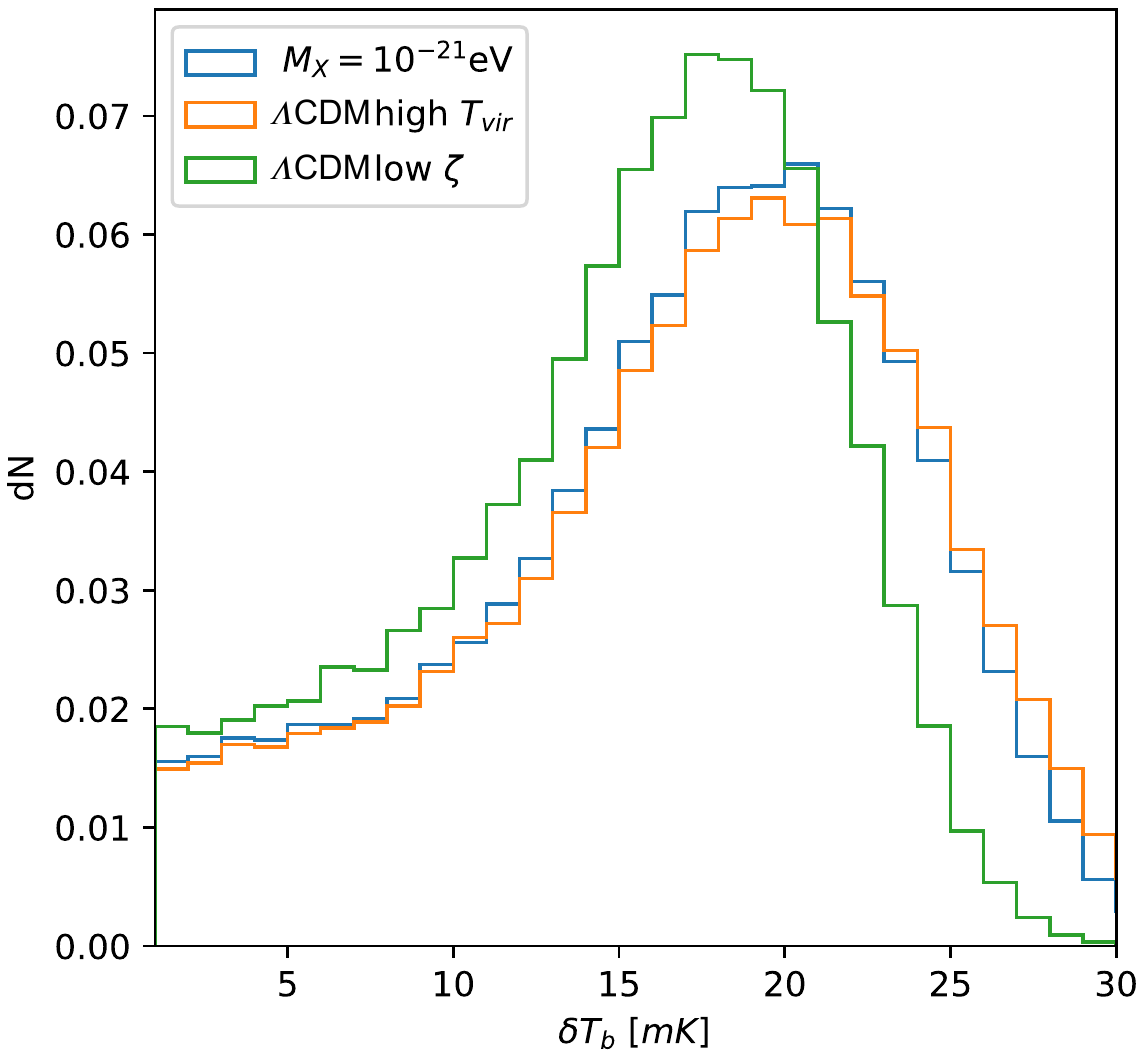}
    \\[-0.5cm]
    \includegraphics[width=0.7\columnwidth]{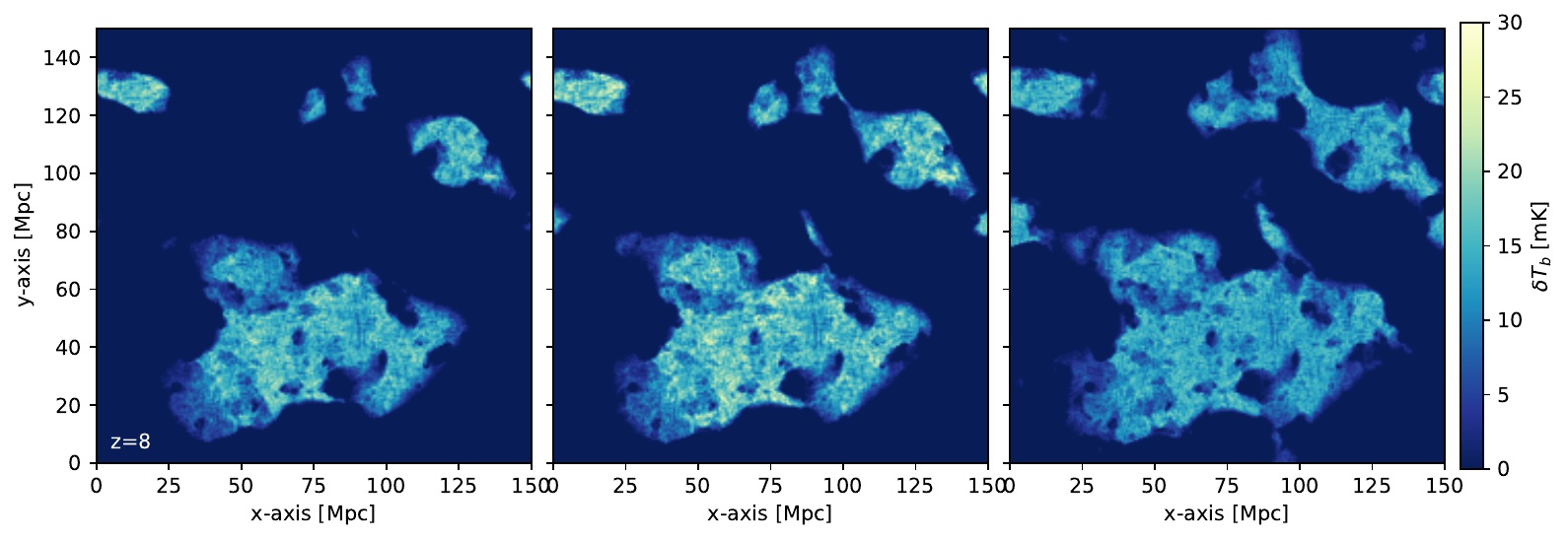}
    \includegraphics[width=0.26\columnwidth]{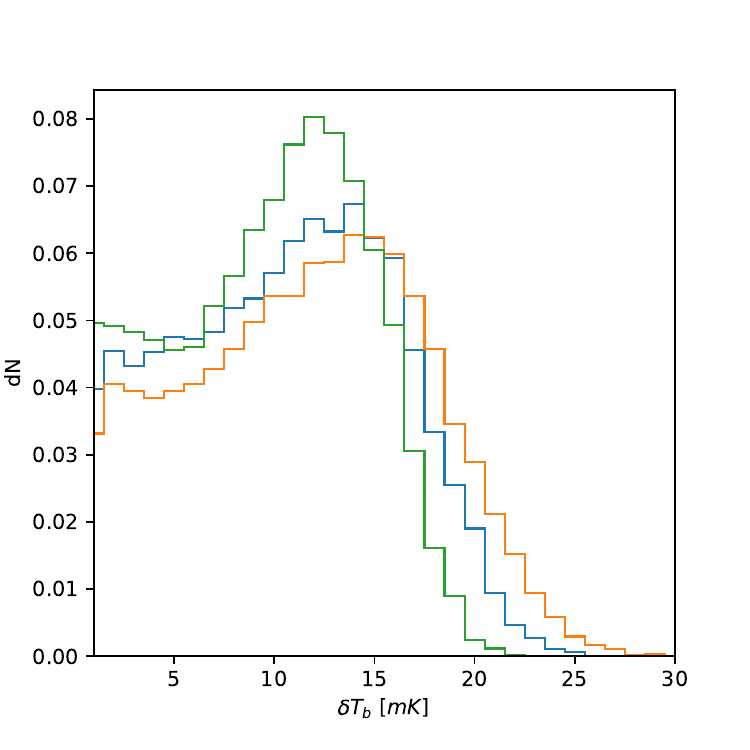}
    \caption{The left three panels we show 2-dimensional slices through the simulation cube for various quantities and redshifts. On the right we show the 1-D histogram of cell values associated with those images. {\em Top: } The differential brightness temperature of the 21cm signal at z=9.5 for i)  an axion DM model with fiducial $T_{vir}$ and $\zeta$ ii) a fiducial $\Lambda$CDM model with a high $T_{vir}=4\times10^4K$ and iii) a fiducial  $\Lambda$CDM model with a low value of $\zeta=15$. These 3 `degenerate' models are the same as those shown in figure~\ref{fig:reionize2}.
    {\em Bottom: }The same as above but now at z=8.0.}
    \label{fig:degeneracy}
\end{figure*}

In figure~\ref{fig:degeneracy} we now take a closer look at the seemingly degenerate models depicted in figure~\ref{fig:reionize2}, i.e i) an axion model with $M_X=10^{-21}eV$ and fiducial astrophysical parameters ii) a fiducial $\Lambda$CDM model with a high value of $T_{vir}=4\times10^4$K and iii) a fiducial $\Lambda$CDM model with a low value of $\zeta=15$. In the top panels of figure~\ref{fig:degeneracy} we show the 2-dimensional spatial distribution of the differential brightness temperature field within a thin slice of $\sim2$cMpc at z=9.5 when $\langle x_H \rangle = 0.5$. The 3 maps are visually similar and difficult to differentiate. If, instead,  we focus on the 1D distribution of pixel values in the right hand panel,  the high $T_{vir}$ $\Lambda$CDM and axion models are almost indistinguishable which may be a cause for concern in our goal to constrain the axion mass. The low $\zeta$ $\Lambda$CDM model shows a deviation from the axion model using this 1$^{st}$ order statistical description of the field. 


Remembering from figure~\ref{fig:reionize2} that the difference in the mean neutral fraction, $\langle x_H \rangle$, between these models has a redshift evolution, we now look at the field at an evolved redshift of z=8 in the bottom panels of figure~\ref{fig:degeneracy}. As we might have expected the different redshift evolutions of these models result in visually different fields. These can be more quantitatively understood from the 1D distributions on the right. 
In both the upper and lower right hand panels of figure~\ref{fig:reionize2}, the high $T_{vir}$ $\Lambda$CDM model peaks at a higher brightness temperature compared to the low $\zeta$ model. 
This is because high $T_{vir}$ preferentially selects heavier halos for the ionization sources, while a low value of $\zeta$ applies a low ionizing efficiency to all halo masses. Thus the axion model is closer to high $T_{vir}$ $\Lambda$CDM because the power suppression wipes out smaller structures that don't reach their $T_{vir}$ threshold.

\begin{figure}
    \centering
    \includegraphics[trim={0cm 2.0cm 0cm 2cm},clip,width=0.7\columnwidth]{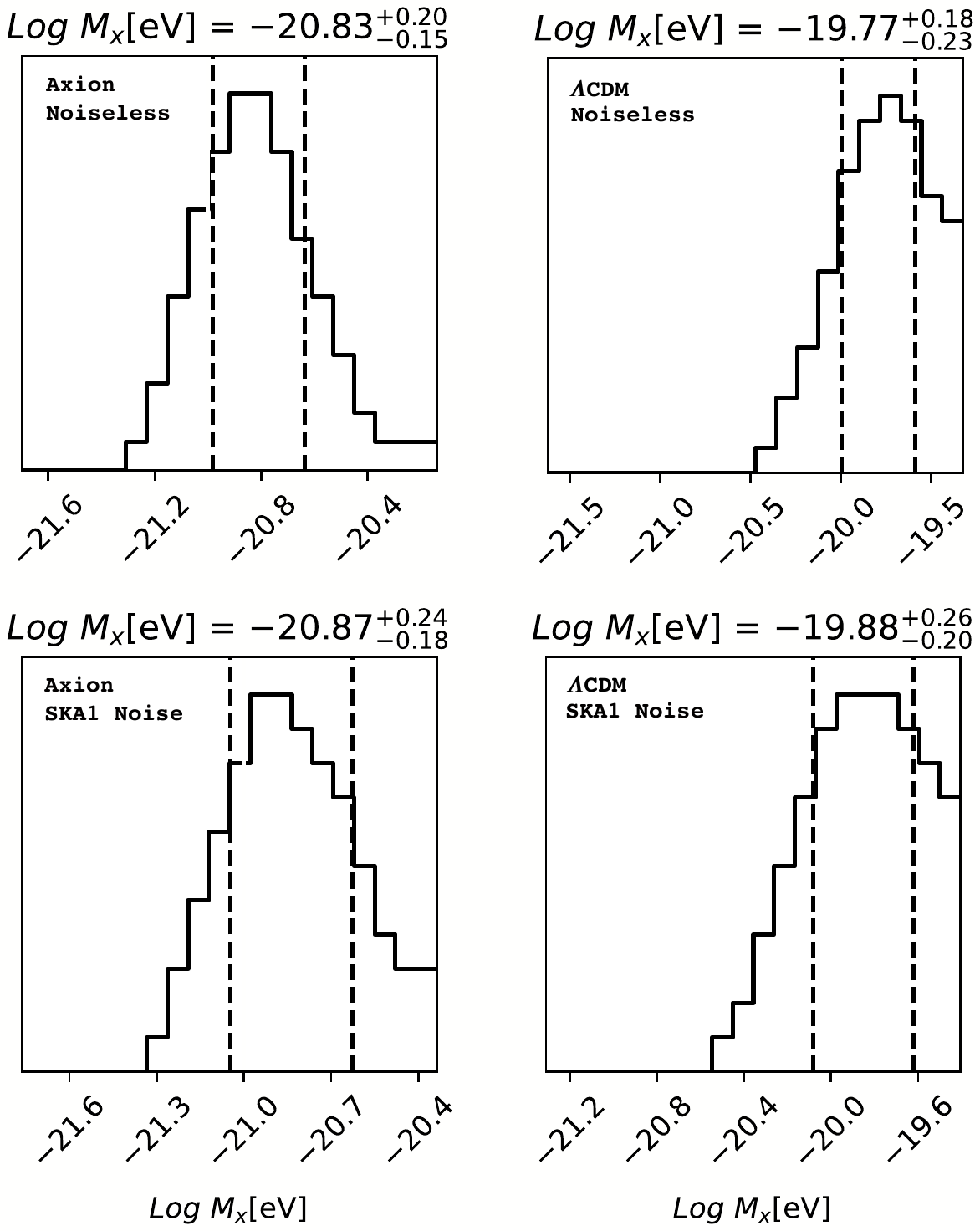}
    \caption{{\em Top Left:} The distribution of our CNN's predictions on 100 noiseless simulations with varying $T_{vir}=10^3 - 10^5$ and $\zeta = 10-60$ and fixed axion mass of $10^{-21}$eV (denoted by the vertical blue line). The vertical dashed lines show the 1-$\sigma$ (68\%) credible interval. {\em Top Right:} Same as the left plot but the simulations are created using $\Lambda$CDM model without axions. {\em Bottom:} As above but now each simulation has a random Gaussian noise realisation added to its signal and it is convolved with a smoothing length applicable to SKA1-Low core design computed through eq.~\ref{eq:resolution}.}
    \label{fig:CNNdist2}
\end{figure}

We proceed, once again, to constrain the axion mass, although we now relax our assumptions regarding the astrophysical modeling and no longer fix the values of $\zeta$ and $T_{vir}$. We thus aim to essentially marginalise over these nuisance parameters rather than constrain them \footnote{See however, for instance, \citet{Kapahtia:2021eok} for the attempts to constrain these astrophysical parameters, given similar observations to those we consider here.}.

We repeat the analysis of the previous section by creating a further 400 simulations that span not only axion mass but also the astrophysical parameters in the range $\zeta = 10-60$ and $T_{vir}=10^3 - 10^5$K. The three model parameters $\{\log M_X, \log T_{vir}, \zeta\}$ were sampled linearly from a latin hypercube design, which has been shown to be an optimal space filling strategy that avoids repeating similar parameter values \citep{eglajs:1977,mackay:1979}.\footnote{Implemented using the \texttt{PYTHON} package \texttt{pyDOE} \url{https://pythonhosted.org/pyDOE/}}

We again train two networks, one with the ideal simulations and one with the SKA1-Low noise simulations. As before we create 100 simulations each for 4 scenarios: a $\Lambda$CDM model and axion DM with $M_X=10^{-21}$eV, both with and without noise+smoothing. These sets of simulations will test the uncertainty of the axion mass prediction within the trained networks.

In the top panels of figure~\ref{fig:CNNdist2} we show the distribution of $M_X$ predictions for the $M_X=10^{-21}$eV simulations on the left and for $\Lambda$CDM on the right, under noiseless/ideal conditions. We find that the recovered axion mass for the $M_X=10^{-21}$eV simulations is $M_X=1.48^{+0.86}_{-0.43}\times10^{-21}$eV, while the $\Lambda$CDM simulations have an inferred axion mass of $M_X=1.70^{+0.87}_{-0.70}\times10^{-21}$eV.

In the bottom panels of figure~\ref{fig:CNNdist2} we show similar results but now under realistic noise conditions parameterized by the Gaussian noise given in eq.~ \ref{eq:noise} and the resolution of eq.~\ref{eq:resolution}. As we might expect the distribution of predicted axion masses is broadened and the inferred uncertainty is increased. In the bottom left we see that the mass of the axion is estimated to be $M_X=1.35^{+0.99}_{-0.46}\times10^{-21}$eV, which again recovers the true input mass. In the bottom right we see that the $\Lambda$CDM simulations are estimated to have an axion mass of $M_X=1.32^{+1.07}_{-0.49}\times10^{-20}$eV. We can infer from this that the axion as the whole DM contribution can be detected using SKA1-Low at 68\% if the axion mass is $M_X<5.25 \times10^{-21}$eV. 

We see that if the astrophysical parameters are treated as nuisance variables their uncertainty does not bias our estimate of the axion mass, however it does degrade the precision to which the axion mass is recovered. This translates to approximately a factor of 2(3) increase in the error of the axion mass for the noiseless (SKA1-Low noisy) simulated images. The tomographic information is of great help in disentangling  axion mass estimates and the astrophysical parameters, and our results are encouraging for the future prospects on the axion search or more generally the search for a model beyond the conventional $\Lambda$CDM through the machine learning. 

\edit{We also tested the sensitivity of the global 21cm signal to the axion mass. Using the same simulations as test data we created a simpler 4 layer dense neural network with 1024 neurons and train it as before. We find that the network can indeed recover the axion mass, however the errors are approximately twice larger than those obtained via the full 3D 21cm field.}

\section{Conclusions} \label{sec:conclusions}

In this work we created simulated maps of the future SKA1-Low 21cm radio survey under an ultra light axion-like dark matter model. We found that the redshift at which the universe becomes half reionized depends strongly on the mass of the axion particle.

We created a set of simulations that sample a range of axion masses and astrophysical parameters. 
Using a machine learning approach, we trained a convolutional neural network to regress from the simulated 3-dimensional differential brightness temperature field directly to the axion mass, essentially marginalizing over the astrophysical parameters. 

We found that the neural network could successfully recover the input truth values of the testing data over a wide range in axion masses with a precision of $\sim20\%$. In order to assess the uncertainty on the recovered value we created a further 100 independent simulations each for an axion model and a  standard CDM model and observed the distribution of their predictions from the pre-trained network. We found that for an axion mass of $M_X=10^{-21}$eV we could recover its mass with an uncertainty of $^{+5.94}_{-4.80}\times10^{-21}$eV.

We conclude that this approach could be used to detect axion dark matter using SKA1-Low at 68\% if the axion mass is  $M_X<1.86 \times10^{-20}$eV although this can decrease to $M_X<5.25 \times10^{-21}$eV if we relax our assumptions on the astrophysical modeling. These results however depend on the assumption of Planck 2015 cosmological parameters and the specific design parameters of the future SKA1-Low configurations.

While this work is encouraging in the search for axion dark matter, there are several important systematics that we did not consider here but leave for future work. These include the foreground contamination from radio emission in our own galaxy, redshift-space distortions and non-linear evolution of the density field. 
 
\acknowledgments{
We thank Jaehong Park for guidance in running 21cmFAST and Sungwook E. Hong, Dongsu Bak, Jaewon Lee and Hyunbae Park for their invaluable comments and suggestions. We also thank the anonymous referee for their kind comments and suggestions that have helped improve this manuscript. 

This article was supported by the computing resources of Urban Big data and AI Institute (UBAI) at the University of Seoul.



This work acknowledges support via the Basic Science Research Program from the National Research Foundation of South Korea (NRF) funded by the Ministry of Education (2018R1A6A1A06024977 and 2020R1I1A1A01073494) and the Institute for Basic Science (IBS-R018-D1). KK thanks Kobayashi-Maskawa institute at Nagoya University for hospitality through JSPS core-to-core program (JPJSCCA20200002) and Grant-in-Aid for Scientific research from the Ministry of
Education, Science, Sports, and Culture (MEXT), Japan (16H06492). JA has received funding from the European Union’s Horizon 2020 research and innovation programe under grant agreement No. 776247 EWC.
}


\bibliography{paper}{}
\bibliographystyle{aasjournal}

\end{document}